# COVID-19 Should be Suppressed by Mixed Constraints from Simulations on Constrained Scale-Free Networks

Yukio Ohsawa

Department of Systems Innovation, School of Engineering, The University of Tokyo
ohsawa@sys.t.u-tokyo.ac.jp

**Abstract.** The spreading of virus infection is here simulated over artificial human networks. Here, the real-space urban life of people is modeled as a scale-free network with constraints. A scale-free network has been adopted for modeling on-line communities so far but is employed here for the aim to represent peoples' social behaviors where the generated communities are restricted reflecting the spatiotemporal constraints in the real life. As a result, three findings and a policy proposal have been obtained. First, the height of the peaks in the time sequence of the number of infection cases tends to get reduced corresponding to the upper bound of the size of groups where all members meet. Second, if we adopt the constraint on $m_0$, the number of all other people one meets separately each at a time, to the range between 2 and 8, its effect on the suppression of infections may be weak as far as we allow group meetings of size $W$ of 8 or larger. Third, such a moderate constraint may temporarily seem to work for the reduction of infections in the early stage but it may turn out to be just a delay of peaks. Based on these results, a policy is proposed here: for quickly suppressing the number of infections, restrict $W$ to less than 4 if the constraint to make $m_0$ at most 1 is too strict. If $W$ is set to less than 4, setting $m_0$ to 4 or less works for quick reduction of infections according to the result.

## 1 Introduction

This study has been motivated by the recent pandemics of COVID-19. So far, the infection spreading has been modeled in a number of studies [1 – 5]. In [1], the dynamics have been modeled to explain the increase of infections in the early stage on the power-law followed by exponential suppression. This model has been used and extended to explain the exponential growth, power-law behavior, and then exponential decline in the death rate [2], and to compare the temporal trends of daily cases in compared countries [3]. Strategies for network manipulation has been also explored, one of which is the control of the spreading dynamics by interacting with features of a network such as infection rate [4], and another is network optimization by removing/rewiring links that is an NP-hard problem [5]. The solutions for these problems include centralized control e.g. to suppress infection rates and a distributed approach where individuals discern the critical in-network contacts. The guidelines for attention and actions obtained from such researches should be considered for designing politics for controlling epidemic spreading.

In the literature of general theories applicable to various influence propagation phenomena on a social network, a lot of findings have been obtained. For example, it has been found that the cascades in skewed human networks are triggered by large degree nodes, whereas small degree nodes may trigger cascade other networks [6]. We can find models for other social phenomena than virus spreading. In [7], the diffusion of innovation has been simulated in artificial social networks and found the delayed propagation to low-degree nodes from high-degree nodes, which is of greater impact if informative and nominal contacts from neighboring nodes are mixed. Although this study is not applied for virus infection events, we find the authors' interest in the mutual influence between high-degree nodes and low-degree nodes in which the role of the latter is coming to be focused. This interest differs from the interest in tail-fat networks or from the detection of low-frequency influential nodes. Further, an encouraging fact is that the authors of [2], [7], and others having been working on network dynamics are contributing to providing points to consider in designing policies for controlling epidemic spreading [8].

The question we address in this paper is simple. Is the spreading suppressed as expected if people are forced to reduce the contacts with others in the network? If so, what should the governmental policy be like? In this paper, we show a quick report based on a simple simulation of infection spreading in a social network of people in an urban environment. In this environment, people get interested in attending a place where a lot of others are attracted. This interest may not be directed to any particular people in the place, but to something that may be an object shown in the place or the atmospheric mood. In this place, a new member may come close to other humans, touch objects implemented in the place, and breathe the air in the place. These behaviors of people are similar to the social dynamics known to form a scale-free network [9] where new nodes get connected to high-degree nodes. However, there are real-space constraints on the dynamics that are 1: the restriction of space (the upper bound on the number of people to meet at a time), and 2: the restriction of time (the number of other people to meet if one may meet just one other at a time). In this paper, we aim to find some clues to discuss the question above by simple simulations using a constrained scale-free network reflecting the two kinds of restrictions.

## 2. The model

### 2.1 Scale free network as the back bone

The established model of the process to generate a scale-free network (SFN) proposed by Barabasi and Albert [9] is described as follows. The process is here described to fit the revision in the next section.

```
1: G := {V, E}
2: V = {node_1,node_2, …, node_m₀}
3: E = {edge_i, j} for all node_i∈ V, node_j∈ V (i ≠j)
4: for i = m₀+1 : N do
5:    add node_ i to V
6:    for j = m₀+1 : N  (i ≠j) do
7:       if degree.RandomRank(j) < m₀
8:          add edge(i, j) to E
9:          degree(i) ← degree(i)+1;
9:          degree(j) ← degree(j)+1
10:      end if
11:   end for
12: end for
```

$V$ and $E$ mean the sets of nodes and edges in the graph respectively, of which the combination is defined to be the graph $G$. The graph $G$ starts from a clique of $m_0$ nodes and incremented by adding one node node_i at each time. Then $m_0$ other nodes are chosen as new link destinations from node_i according to the probabilities given in proportion to the node degrees. Here $m_0$ is the initial degree of each node (the number of edges connected to a new node), and RandomRank($j$) means the rank (0, 1, 2, …) of node_j in choosing on the probability given in proportion to its degree denoted degree($j$). That is, the values of degree($j$)*random(0,1) for $j$'s get compared for ranking of which the top $m_0$ are chosen. In the real human society, $m_0$ can be regarded as the number of touchable people i.e., all other people one can meet separately if preferable, not necessarily in a group meeting.

It is well known that the distribution of the degree of nodes follows the power-law, that means a few nodes in $V$ occupy a large portion of the edges in $E$. Networks such as WWW and online communities [10], cellular networks in biology [11], patients of sexually transmitted diseases [12], etc. have been known to for SFN following the power-law degree distribution. We basically regard the generation process above of SFN as a model capturing real social behaviors of people approximately, comparing with other existing models. For example, all the nodes are connected to the same number of edges in a regular graph, which is inconsistent with the real inequality of peoples' social activities in the real world. The small-world networks on the WS (Watts-Strgats) model start from a regular graph and move a certain number of randomly selected edges connected from each node, that we humans do not do in the daily activities [13]. In the more recently proposed Mediation-Driven Attachment Model [14, 15], each new node first picks an existing node at random and connects not directly with this but with *a certain number* of its neighbors also picked at random. This is an extension of SFN supposed to result in having each new node connected to "rich" people linked to a large number of "poor" people of low degrees by choosing link destination nodes by the probability estimated to be inverse of the harmonic mean (IHM). Although this may make the network fit to the real society, we do not choose as the model for this paper because the attendants of places do not have a bias to low-degree nodes in the daily life human behaviors. Thus, considering the suitability of each model to the intuitive understanding of the social life of people, we take SFN as the backbone and revise it with introducing constraints corresponding to the spatiotemporal restrictions in the real life of people as in the next section. Reference [12] above encourages this choice in the sense a virus infection network has been shown to follow the power-law distribution.

**2.2 Constrained on a scale-free network reflecting spaciotemporal restrictions**

The SFN may be, as discussed above, regarded as a natural model for capturing human social behaviors as far as there is no restriction on the reach of social behaviors of each human. On the other hand, the physical constraints in the real living environment of humans restrict the width of the room where people may gather, and the time is also restricted so that a person cannot meet as many people one likes to meet. Thus, the scale-free network cannot be used as it is to model real-space social behaviors of people. In this sense, let us revise the process above as follows.

An important conceptualization here is that not only humans but also rooms/spaces of a place or things are dealt with as nodes here assuming these are working as entities attracting people, and these entities are dealt with to meet the requirements of humans who have strong influences on others. Furthermore, the virus is wrapped in saliva and cast into the air in meeting places. We reflect the spatiotemporal restrictions as constraints in *1, *2, and *3 below. In the lines 3, 11, and 12 below (*1 and *2), at each time a new node is added, the degree of the link destination node (the partner of the new connection) from the new node is constrained by the given upper bound of $W$. If the node comes to be more popular than having $W$ links, the node is removed as in line 12. This means the room or space of capacity of $W$ or larger is forbidden as in the protection of COVID-19 infection spreading. However, to add allowed groups of up to the capacity of $W$ attendants by a certain frequency considering the load of the group-meeting organizer, we add the lines 16 and 17 where some nodes get larger preference than others when a new node explores link destinations (*3). $\alpha$ is a constant set to 5. The use of $W$ in this line is not necessary (any sufficiently large value satisfies the requirement here) because the degree of the destination node ($j$ here) is finally bounded by $W$ after all due to *2.

```
1: G := {V, E}
2: V = {node_1,node_2, ..., node_m₀}
3: E = {edge_i, j} for all node_i∈ V, node_j∈ V (i ≠j)                    *1
4: for i = m₀+1 : N do
5:    add node_i to V
6:    for j = m₀+1 : N  (i ≠j) do
7:       if degree.RandomRank(j) < m₀
8:          add edge(i, j) to E
9:          degree(i) ← degree(i)+1;
10:         degree(j) ← degree(j)+1
11:         if degree(j) > W
12:            remove j and all edges connected to j from E                *2
13:         end if
14:      end if
15:   end for
16:   if mod(i, αW) = 0
17:      add W to degree(i)                                                *3
18:   end if
19: end for
```

## 2.3 The model of infection spreading dynamics

For each node in the network generated above, the three-step dynamics is considered here for spreading the infection. The first is to catch a virus from neighboring nodes (i.e., humans or things in the place), the second is to be infected, and the third is to be an infector to other nodes. These steps are supposed here to occur as follows, week by week:

*To use the edges to contact neighbors*: the edges in $E$ added by randomly chosen 3% of the steps marked with an asterisk (*2, *3: each step with an asterisk is regarded as a human's or a human group's social behavior at a time.) in the procedure of 2.2 above is activated, corresponding to peoples' activities in the real life. For the event of *1 and *3, the nodes joining the event form a clique and each node catches the virus from the partner if the partner has, following the rule to catch the virus below. For event *2, on the other hand, a new node contacts each of $m_0$ link destinations one by one so that the pair-wise casting and catching of virus may occur between linked two nodes at each time.

*To catch the virus from neighbors*: catch($i$) takes the maximum value of infector(node_$j$) of all node_$j$'s adjacent to node_$i$. In other words, a node catches virus if any neighbor i.e., any adjacent node, has become an infector following the rule below.

*To be infected*: node_$i$ is infected if catch($i$) is larger than a random value between 0 and 1 by uniform probability. This means one gets infected by 50% in the case of a sufficient catch of viruses. The strength of infection is here is given as a real value infected($i$) rather than a discrete judgment considering, the uncertainty of the PCR test. The infection of a node is assumed to fade by a constant r (set to 0.7 corresponding to the recovery in 4 weeks) each week. We assume the recovery occurs by virus's inactivation or the acquisition of immunity

*To be an infector*: If one is infected, one is expected to infect others by the probability of 0.2 here, according to the announcement of WHO on April 10, 2020.

Anyone linked to 100 other nodes is expected to infect others to generate 1.2 other infected nodes in 4 weeks (100 * 0.03 * 0.5 *0.2 *4). This is less than the reported number of infection reproductions by the government [16], but increases with the coefficient corresponding to the number of infected people to meet in one's circumstance. The value infector($i$) is here is also given as a real value and fades by *r* per week.

## 3. Results

Following the procedure above, we made the experimental simulation for 20 trials of 100 weeks/trial, setting the starting infection from a given set of nodes as the 0-th week and the number of nodes ($N$) in $G$ to 1000 and 10000. Let us first show the overall tendencies obtained, then go into the time series of daily new cases for exemplified settings of $W$ and $m_0$. We did these experiments fixing two starting infected nodes, one in a cluster (meaning to have degree $W$) and another out of any cluster.

### 3.1 The observed tendencies of infections under the constraints on $W$ and $m_0$,

As in Table 1, we first find two tendencies in the range of large $W$ and $m_0$ shown by the shadowed part of the table. First, $W$, the upper bound of the allowed size of group meetings (where participants all meet at once), depends positively on the number of infection cases according to the values from left to right cells in each row. Second, the smaller value of $m_0$, which means the number of other people each human meets (which may be in separate meetings), is not responded by the smaller number of new infection cases for the middle-sized or the larger $W$ ($W \geq 10$) according to the values in each column. For $10 \leq W \leq 50$, the largest number of infections occur for $4 \leq m0 \leq 16$ rather than $m_0 \geq 32$. The value of $m_0$ corresponding to the largest number of new infection cases tends to be the smaller for the smaller $W$, as the bold letters in Table 1 show. On the other hand, in the white cells where $W < 8$ or $m_0 < 4$ (especially for less than 2), the number of new infection cases are substantially smaller than in the shadowed part.

In Table 2, the average timing of infection for all the cases is shown. These values mean an expectation about when the infection tends to occur, setting the first infections in this network as the 0-th week. In the shadowed range of this table, the third tendency is observed: the average infection week tends to be the later for the smaller $m_0$. In other words, the number of infection cases for a small $m_0$ is small for the same period as when the upward trend is apparent for larger $m_0$ and then increases after the period.

Table 1. The average number of infection cases per week for 100 weeks (columns for $W$: the upper bound of group size, rows for $m_0$: the number of touchable people for each human). The bold letters show the two largest values in each column, the underlines the two in each row. The values are shown to the first decimal place just for the space limit but are compared including the second in this ranking.

$N = 1000$

|  | $W$=100 | 50 | 20 | 10 | 4 | 3 | 2 | 1 |
|---|---|---|---|---|---|---|---|---|
| $m_0$=32 | **<u>49.9</u>** | <u>20.7</u> | 1.5 | .8 | **.5** | **0.4** | .1 | .0 |
| 16 | **<u>47.3</u>** | **<u>47.2</u>** | 5.8 | .4 | **.2** | **0.2** | **.1** | **.1** |
| 8 | <u>43.7</u> | **<u>44.2</u>** | **33.8** | 1.7 | .1 | .1 | **.1** | **.1** |
| 4 | <u>26.8</u> | <u>26.4</u> | **14.1** | **4.8** | .1 | .1 | .0 | .0 |
| 2 | <u>11.1</u> | <u>4.6</u> | .6 | .3 | .2 | .1 | .0 | .0 |
| 1 | <u>.1</u> | <u>0.1</u> | 0.0 | .1 | .0 | .0 | .0 | .0 |

$N = 10000$

|  | $W$=100 | 50 | 20 | 10 | 4 | 3 | 2 | 1 |
|---|---|---|---|---|---|---|---|---|
| $m_0$=32 | **<u>474.9</u>** | <u>17.9</u> | 1.4 | 1.0 | **.5** | **.4** | **.5** | **.3** |
| 16 | **<u>498.6</u>** | **<u>495.9</u>** | 16.0 | .4 | **.3** | **.2** | **.2** | **.2** |
| 8 | <u>411.3</u> | **<u>497.2</u>** | **443.9** | 2.3 | .1 | .1 | .1 | .1 |
| 4 | <u>410.9</u> | <u>474.8</u> | **378.5** | **191.6** | .0 | .1 | .1 | .1 |
| 2 | <u>144.9</u> | <u>133.9</u> | 41.3 | **4.6** | .2 | .1 | .0 | .0 |
| 1 | <u>0.1</u> | 0.0 | 0.1 | .0 | <u>.5</u> | .0 | .0 | .0 |

**Table 2. The average timing, i.e., the average number of weeks after the beginning of the simulated spreading, of all infection cases**

| $N = 1000$ | | | | | | | | |
|---|---|---|---|---|---|---|---|---|
| | W=100 | 50 | 20 | 10 | 4 | 3 | 2 | 1 |
| $m_0$=32 | 12.6 | 14.6 | 15.3 | 13.7 | **12.2** | **10.1** | **9.6** | **8.1** |
| 16 | 14.4 | 17.5 | 16.1 | 12.5 | 11.0 | **10.8** | **8.2** | 6.8 |
| 8 | 18.4 | 21.1 | 24.1 | **28.8** | 8.24 | 6.8 | 8.0 | 6.5 |
| 4 | **23.5** | **36.7** | **38.4** | **52.1** | 8.6 | 6.7 | 5.8 | 6.7 |
| 2 | **31.1** | **46.4** | **54.1** | 21.8 | **11.8** | 7.8 | 7.0 | 6.6 |
| 1 | 14.1 | 13.3 | 8.4 | 7 | 8.1 | 4.1 | 4.3 | **7.0** |
| $N = 10000$ | | | | | | | | |
| | W=100 | 50 | 20 | 10 | 4 | 3 | 2 | 1 |
| $m_0$=32 | 7.1 | 15.1 | 15.9 | 14.5 | **11.3** | **11.2** | **10.0** | **8.1** |
| 16 | 10.2 | 11.2 | 45.8 | 10.9 | 7.9 | 8.5 | 7.2 | 6.8 |
| 8 | 14.2 | 22.8 | 39.4 | 31.7 | 9.8 | 6.9 | **9.5** | 6.5 |
| 4 | **36.9** | **36.4** | **48.7** | **74.2** | 6.2 | 8.7 | 3.7 | 6.7 |
| 2 | **53.6** | **63.7** | **67.1** | 51.8 | 10.5 | 11.2 | 8.3 | 6.6 |
| 1 | 17.7 | 17.0 | 9.9 | 8.6 | 5.6 | 7.8 | 5.0 | **7.0** |

### 3.2 The time series of new infection cases

This section aims to show some pieces of evidence to support the tendencies above. Figures 1 through 3 show the average number of new infection cases per week in each condition for 20 trials of simulation. Note each curve thus shown here shows not a history in one trial but the mixture of the sequences of the averaged trials among which the periods of steep increase differ. It is thus meaningless to measure the declination of the curves at each week or to fit the curve to exponential or power-law as in [2] or [3]. The purpose here is to see the overall tendencies such as an early or a slow increase in the larger time scale such as months and the changes in the number of infection cases in a similar time scale.

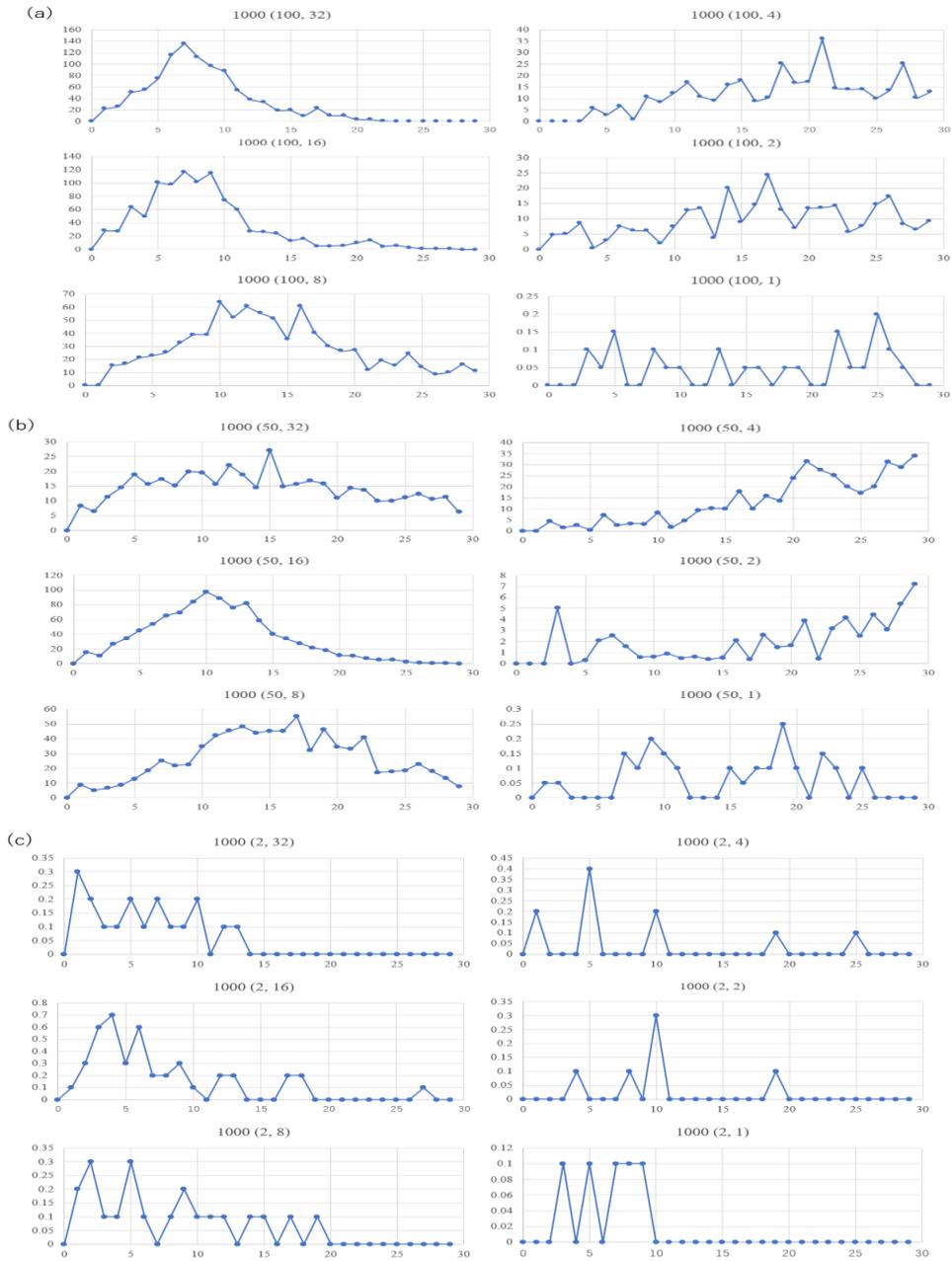

**Figure 1. The number of new infection cases for varying widths of constrained SFNs of 1000 nodes.** $X$ ($Y$, $Z$) for each curve represents $X$: the number of nodes ($N$), $Y$: the upper-bound of the size of group meetings ($W$), $Z$: the upper-bound of the number of connected others for each node ($m_0$). (a) (b) and (c) shows curve-sets for different Y.

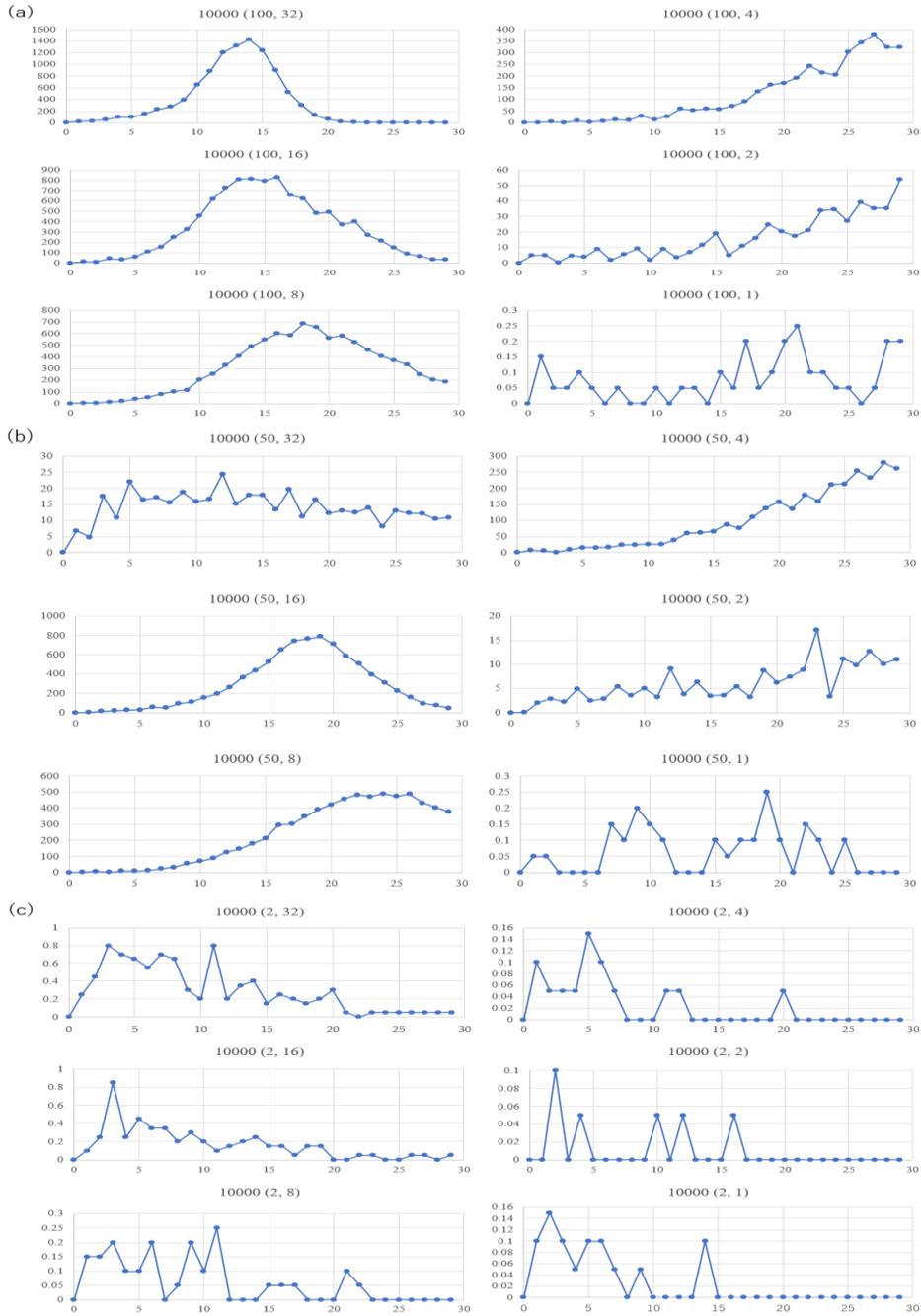

**Figure 2. The number of new infection cases for varying widths of constrained SFNs for 10000 nodes. *X* (*Y*, *Z*) means the same as in Figure 1.**

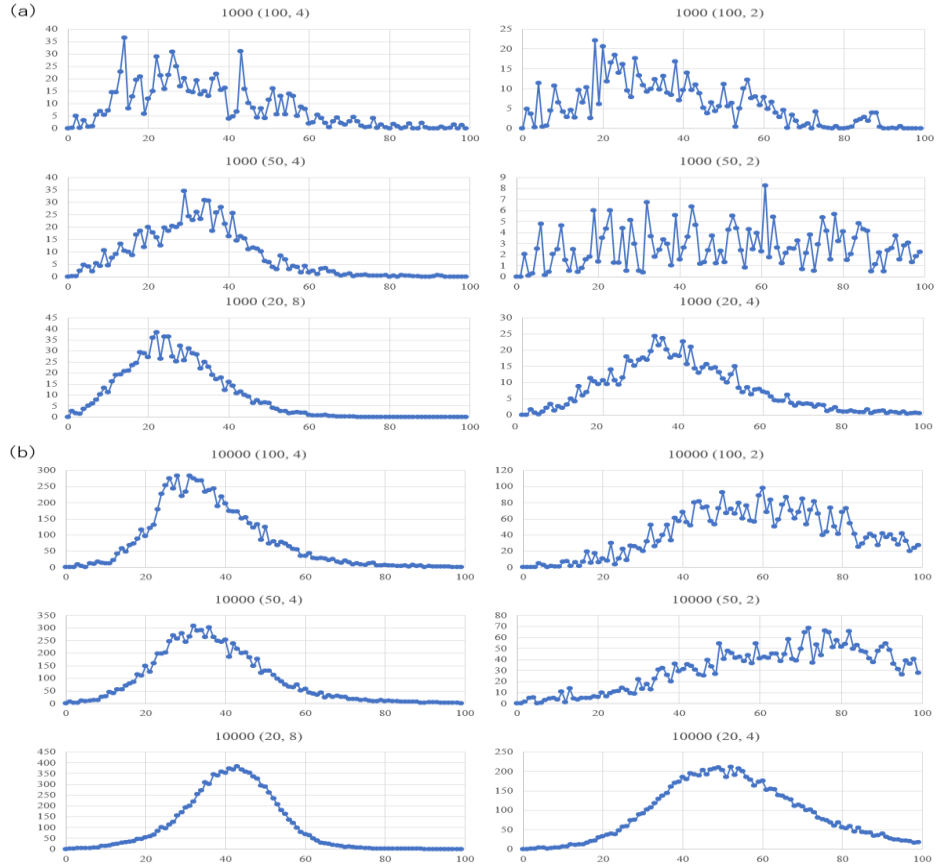

**Figure 3.** The number of new infection cases for varying widths of constrained SFNs of (a) 1000 nodes and (b) 10000 nodes, for a longer time range for the curves that showed up-ward tendencies for 30 weeks. The curves of $W = 100$ and 50 are the extensions of corresponding curves in Figures 2 and 3. It is found here the peaks are delayed in comparison with larger $m_0$.

## 4. Discussions

The first finding above that $W$ depends positively on the number of infection cases suggests the largest allowed number of participants in a meeting should be restricted for suppressing the number of infected cases. This suggests the policy to forbid face-to-face group meetings (i.e., $W \geq 10$) is expected to work for suppressing infections spreading.

The second finding is the tendency that the value of $m_0$ causing the worst effect, i.e., the largest number of infection cases, tends to take a smaller value than the largest value taken in the experiment if $W$ is large. The author's hypothesis for explaining this surprising result is that the people one meets tend not to have acquired immunity yet if the number of people to meet is restricted to a narrow community. This lack in immunity is supposed to occur in the sparsely connected parts of the entire graph $G$ because people on the nodes in such parts are of lower degree and fewer opportunities to touch others. If there are large meeting groups in such a case, the hypothesis goes that the infections in these groups attack the nonimmune part of the network. However, if $m_0$ is constrained further to 1, according to Table 1, the risk is reduced because the casting and catching of viruses occur in lower probability. Thus, we should allow people to meet only one restricted member i.e., set $m_0$ to 1, as far as we allow group meetings.

The above two findings can be reviewed referring to Figure 4 where the results of other conditions than above are added (fourteen values of $W$ in $1 \leq W \leq 100$ and six values of $m_0$ in $1 \leq m0 \leq 32$). Here, $t_d$ means the start of the final downtrend of new infection cases in the time series, given by $\min(\tau \mid mean_{t \in [\tau-r,\tau]} newcases(t) > mean_{t \in [\tau,\tau\mp r]} newcases$ and $\forall t > \tau \{mean_{t \in [\tau-r,\tau]} newcases(t) \geq mean_{t \in [\tau,\tau\mp r]} newcases(t)\})$. $r$ represents the length of the periods before and after the $\tau$-th week referred to for evaluating the trend (down or up). Comparing the left ($N$ = 1000) and the right ($N$ = 10000) of Figure 4, the contours segmenting different values of $t_d$ do not differ substantially for different sizes of the network. The first finding above is refined based on these contours. That is, zone A in Figure 4 meaning the quickest suppression of new infection cases for both $N$ = 1000 and 10000, is given by restricting both $W$ and $m_0$ to the low-value ranges. Thus, the restriction of $W$ in the first finding is a necessary but not a sufficient condition for the quick suppression. In addition, the second finding above is now extended to the dimension of time as well as the sheer number of infections in Section 3. That is, in the boundary between zones B and C for $W$ = 8, $t_d$ comes to be larger in the range of moderately restricted $m_0$ (e.g., $2 \leq m_0 \leq 4$) than for $m_0$ of a larger value. This tendency is found in the similar range of $W$ and $m_0$ to the second finding in Section 3 where the number of infection cases was simulated, so we can assume that a policy of restriction on $W$ and $m_0$ that reduces the number of infections can also quicken the reduction. The zone D in the dotted circle is not meaningful for the discussion here because this zone shows the saturation of new cases reaching a large portion of the entire population.

The third finding in Section 3 was about the timing of infections. In Figure 1 ($N$ = 1000) and Figure 2 ($N$ = 10000), we obtained the up-ward trend in the first five to ten weeks that is stronger for the larger $m_0$ (note: the vertical axes has different scales) which seems to disappear for the same period if $m_0$ is controlled to a smaller value. A moderate constraint where $m_0$ ranges in $2 \leq m_0 \leq 8$ may temporarily seem to work for the reduction of infections in the early stage within 30 weeks from the beginning. However, this does not mean the uptrend really disappears but may be just delayed to a later period. As in Figure 1 and 2, the increase in the number of new infection cases for smaller $m_0$ continues for all the 30 weeks, which is a period that covers the entire hills (i.e., both the uptrend and the downtrend) of new infection cases for the larger $m_0$. In Figure 3, showing 100 weeks in the conditions of smaller $m_0$, we find the slow increase for the first 30 weeks in each condition shown in Figures 1 and 2 was really a part of a long-lasting increase. In Figure 3, where $N$ ($W$, $m_0$) is 10000 (50, 2), the increase is found to last for 80 weeks i.e., 20 months. The peaks of new infection cases tend to shift to the later weeks for smaller $m_0$ in the range of large $W$. This trend is consistent with the results in Fig 4, where we find the delayed reduction for $2 \leq m_0 \leq 8$ in the range of large $W$ ($W \geq 8$).

From the discussion above, in summary, the following policies are recommendable

(1) For suppressing the number of infections, restrict $W$ to less than 10, or $m_0$ to 1.
(2) For suppressing the infections quickly, restrict $W$ to less than 4, or $m_0$ to 1 (zone A in Figure 4). Coupled with (1), this is our recommendation for suppressing infections both in quantity and quickness. If the latter meaning to meet only one other person, is a too hard constraint, $W$ should be restricted to less than 4. If $W$ is restricted so, setting $m_0$ to 4 or less can further quicken the reduction (zone A in Figure 4).
(3) A moderate restriction on the widths of contacts ($W$ or $m_0$) may seem to appear as a reduction of infections for a period, but we should take care of the real peak of infections that may come later.

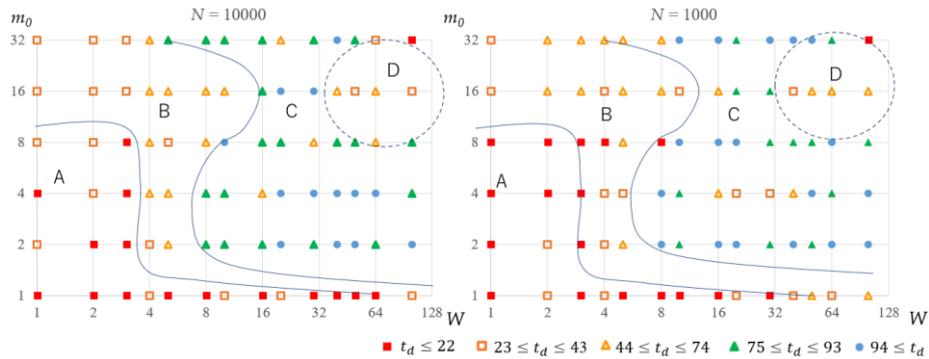

**Figure 4. The ranges of ($W$, $m_0$) corresponding to the ranges of $t_d$, the week of the first $r$-weeks downtrend of the last hill of the curve. The zones A, B, and C respectively mean where $t_d$ is within 22 weeks (ca 5 months), 23 to 74 weeks, and longer. Zone D means the exceptions of A and B where $t_d$ is within 74 weeks but $W$ and $m_0$ are in the range of largest values.**

## 5. Conclusions

The infection spreading has been simulated over human networks where the constraints are given by the maximum size of a group in which people meet at once ($W$) and the maximum number of all other people one can meet separately if preferable ($m_0$). The real-space urban life of people is modeled as a scale-free network with constraints on the constants on the two width factors $W$ and $m_0$ above. As a result, three findings have been obtained as in the discussion, that can be put within this one paragraph. That is, for quickly suppressing the number of infections, we recommend a policy to restrict $W$ to less than 4 and $m_0$ to 4 or less, if it is too hard to set $m_0$ to 1 meaning each person can meet only one other person for all the period of a virus infection crisis. Otherwise, i.e. if the politics uses a weaker restriction on the widths of contacts, we should take care that the infection may not be suppressed significantly or the real peak of infections may come later.

The results in this paper might be associated with the assertion that there is a risk of infection spreading in a society where people in each household decide to maintain an in-person social connection with one person from households [17]. However, the suggested policy is not so strict in that we can release the constraint on one-to-one meetings to allow $m_0 = 4$ if people stop all face-to-face group meetings of four or more participants.

## Acknowledgement

This study is in progress, partially supported by grant JSPS Kakenhi 19H05577.